\documentclass[preprints,article,accept,pdftex,moreauthors]{Definitions/mdpi} 
\usepackage{aas_macros}
\usepackage{mathtools}
\usepackage{amssymb,amsmath}
\usepackage{graphbox}
\usepackage{amsmath}
\usepackage{natbib}

\def\lsim{\mathrel{\raise.3ex\hbox{$<$\kern-.75em\lower1ex\hbox{$\sim$}}}}
\def\gsim{\mathrel{\raise.3ex\hbox{$>$\kern-.75em\lower1ex\hbox{$\sim$}}}}

\firstpage{1} 
\pubvolume{1}
\issuenum{1}
\articlenumber{0}
\pubyear{2022}
\copyrightyear{2022}
\datereceived{Nov. 4, 2022} 
\dateaccepted{TBD} 
\datepublished{TBD} 
\hreflink{https://doi.org/} 

\Title{How Spatially Resolved Polarimetry Informs Black Hole Accretion Flow Models}

\TitleCitation{How Spatially Resolved Polarimetry Informs Black Hole Accretion Flow Models}

\Author{Angelo Ricarte $^{1,2}$*\orcidA{}, Michael D.~Johnson $^{1,2}$\orcidB{}, Yuri Y.~Kovalev $^{3,4,5}$\orcidC{}, Daniel C. M. Palumbo$^{1,2}$\orcidD, Razieh Emami$^{1}$\orcidE}

\AuthorNames{Angelo Ricarte, Michael D.~Johnson, Yuri Y.~Kovalev, Daniel Palumbo, and Razieh Emami}

\AuthorCitation{Ricarte et al.}

\address{%
$^1$ \quad Center for Astrophysics | Harvard \& Smithsonian, 60 Garden Street, Cambridge, MA 02138, USA \\
$^2$ \quad Black Hole Initiative, 20 Garden Street, Cambridge, MA 02138, USA \\
$^3$ \quad Max-Planck-Institut f\"ur Radioastronomie, Auf dem H\"ugel 69, 53121 Bonn, Germany\\
$^4$ \quad Lebedev Physical Institute of the Russian Academy of Sciences, Leninsky prospekt 53, 119991 Moscow, Russia\\
$^5$ \quad Moscow Institute of Physics and Technology, Institutsky per. 9, Dolgoprudny 141700, Russia
}

\corres{Correspondence: angelo.ricarte@cfa.harvard.edu}

\abstract{The Event Horizon Telescope (EHT) Collaboration has successfully produced images of two supermassive black holes, enabling novel tests of black holes and their accretion flows on horizon scales.  The EHT has so far published total intensity and linear polarization images, while upcoming images may include circular polarization, rotation measure, and spectral index, each of which reveals different aspects of the plasma and space-time.  The next-generation EHT (ngEHT) will greatly enhance these studies through wider recorded bandwidths and additional stations, leading to greater signal-to-noise, orders of magnitude improvement in dynamic range, multi-frequency observations, and horizon-scale movies. In this paper, we review how each of these different observables informs us about the underlying properties of the plasma and the spacetime, and we discuss why polarimetric studies are well-suited to measurements with sparse, long-baseline coverage.}


\begin{document}

\thispagestyle{empty}

\section{Simulating Black Hole Accretion Flows}
\label{sec:introduction}

The Event Horizon Telescope (EHT) collaboration has produced the first images of supermassive black holes (SMBHs), ushering in a new era of spatially resolved astrophysics at the event horizon \citep{EHTC+2019a,EHTC+2019b,EHTC+2019c,EHTC+2019d,EHTC+2019e,EHTC+2019f,EHTC+2021a,EHTC+2021b,EHTC+2022a,EHTC+2022b,EHTC+2022c,EHTC+2022d,EHTC+2022e,EHTC+2022f}.  The images have been very constraining for general relativistic magnetohydrodynamics (GRMHD) models, which evolve plasma in a Kerr spacetime under the assumptions of ideal MHD.  EHT science has focused mainly on constraining three free parameters: spin, the magnetic field state, and $R_\mathrm{high}$, which is related to the ion-to-electron temperature ratio \citep{EHTC+2019e,EHTC+2021b,EHTC+2022e}.  The SMBH spin, which we will denote as $a_\bullet$, is the dimensionless angular momentum of a SMBH described by a Kerr metric that can vary between $|a_\bullet| \in [0,1)$.  A SMBH's spin reflects its recent assembly history and affects its accretion and feedback processes \citep[see][]{white_paper_spin}.  Meanwhile, the accretion flow's magnetic field structure may vary between ``MAD'' and ``SANE'' states.  In a Magnetically Arrested Disk (MAD), the magnetic flux at the horizon saturates, and the magnetic fields grow dynamically important, resulting in azimuthal asymmetries include flux eruption events \citep{Bisnovatyi-Kogan&Ruzmaikin1974,Igumenshchev+2003,Narayan+2003}.  This contrasts with ``Standard and Normal Evolution'' (SANE), where the magnetic fields remain turbulent and dynamically unimportant \citep{Narayan+2012,Sadowski+2013}.  Finally, the ratio of ion to electron temperature in different regions is highly uncertain, since the mean free path of particles is much larger than the size of the system, and ions are heated more efficiently \citep{Shapiro+1976,Rees+1982,Narayan+1995}.  EHT studies have encapsulated this uncertainty with the post-processing parameters $R_\mathrm{low}$ and $R_\mathrm{high}$, which describe the asymptotic ion to electron temperature ratio at low and high plasma $\beta$ respectively \citep{Moscibrodzka+2016}.  Less thoroughly studied parameters include the electron distribution function (eDF) \citep{Ozel+2000,Mao+2017,Cruz-Osorio+2022,Fromm+2022,EHTC+2022e}, the detailed particle composition of the plasma \citep{Anantua+2020,Emami+2021,Wong&Gammie2022}, and the tilt of a potentially misaligned disk \citep{Fragile+2007,Liska+2021}, which are the subject of many recent and ongoing studies.

\begin{figure*}
  \centering
  \includegraphics[width=\textwidth]{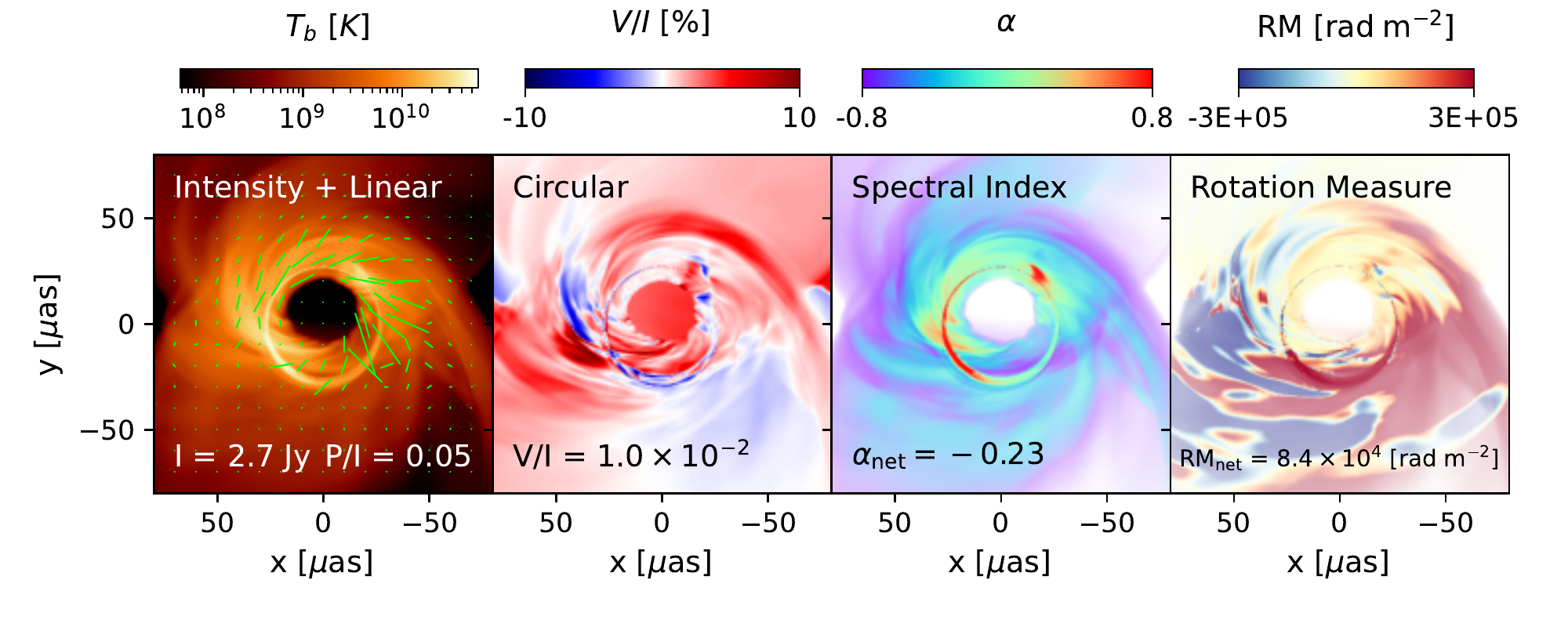}
  \caption{A single GRMHD snapshot ray-traced and scaled to Sgr~A$^*$ properties, with three decades in dynamic range shown.  So far, total intensity maps have been produced for both Sgr~A$^*$ and M87$^*$, a linear polarization map has been produced for M87$^*$, and the remaining observables have yet to be generated for either source.  In the era of ngEHT, we will have access to each of these observables with improved dynamic range and time-domain information, which will greatly inform models of the black hole accretion flow.}  \label{fig:example}
\end{figure*}

In this paper, we review how properties of the spacetime and the plasma become imprinted onto multifrequency polarimetric observables accessible to the EHT and ngEHT.  In \autoref{fig:example}, we plot a single GRMHD snapshot, ray-traced at 214 and 228 GHz with {\sc ipole} \citep{Moscibrodzka&Gammie2018} and scaled to Sagittarius A*.  This particular model is a MAD model with $a_\bullet=0$, $R_\mathrm{high}=40$, and a viewing angle of $50^\circ$.  So far, the EHT has produced total intensity maps for both M87$^*$ and Sgr~A$^*$ and a linear polarization map of M87$^*$.  The remaining maps have yet to be generated for EHT data, and will be explained in detail in this paper.  In brief, circular polarization may arise from both Faraday conversion and intrinsically emitted synchrotron \citep[e.g.,][]{Wardle&Homan2003}, and is especially sensitive to details of the underlying magnetic field geometry \citep{Moscibrodzka+2021,Ricarte+2021,Tsunetoe+2021} and plasma composition \citep{Wardle+1998,Anantua+2020,Emami+2021}.  Meanwhile, the spectral index is the logarithmic derivative of the flux or intensity with respect to frequency, $\alpha \equiv d\log I/d\log \nu$, which helps break degeneracies between number density, temperature, and magnetic field strength \citep{Bower+2019,Ricarte+2022a}.  Finally, rotation measure is the derivative of the electric vector position angle (EVPA or $\chi$) with respect to observing wavelength squared $\mathrm{RM} \equiv d\chi/d\lambda^2$, which encodes Faraday rotation.  Since colder electrons Faraday rotate more efficiently, the RM offers a glimpse into a colder population of electrons, which may exist at large number densities but may be too cold to contribute significantly to the intensity \citep{Moscibrodzka+2017,Jimenez-Rosales&Dexter2018,Ricarte+2020}.  As discussed by the other contributions to this special issue, the ngEHT will enable access to this expanse of information with higher image dynamic range than currently published EHT results (increasing from ${\sim}10$ to ${\gsim}10^3$), improved spatial resolution (decreasing from about 20 $\mu$as to $\sim$10-15 $\mu$as), and time-resolved images of the dynamical activity in both M87$^*$ and Sgr~A$^*$ over hundreds-to-thousands of gravitational timescales. This will result in movies of both the accretion disks and relativistic jets near SMBHs, and here we discuss what physical information each of these maps carry.

\section{Total Intensity and Spectral Index}
\label{sec:total_intensity}

In the millimeter, we observe Sgr~A$^*$ and M87$^*$ near the peak of emission from synchrotron radiation, where the flow transitions from optically thick to optically thin.  Here, the emissivity scales approximately as $j_\nu \propto n B^2\Theta_e^{5/2}$, where $n$ is the electron number density, $B$ is the magnetic field strength, and $\Theta_e$ is the electron temperature (in units of the electron rest mass energy).  Each of these quantities can vary by orders of magnitude among different models, and thus even the total flux is informative for jointly constraining these parameters.  To match the total flux of a given system, the fluid in ideal GRMHD simulations can be rescaled via $n \to \mathcal{M}n$, $B \to \sqrt{\mathcal{M}}B$, and $u \to \mathcal{M}u$, where $u$ is the internal energy and $\mathcal{M}$ is a scalar.  After doing so, both MAD and SANE simulations are capable of matching the total flux of EHT sources at a single frequency, as well as broad image characteristics such as the image size \citep{EHTC+2019e,EHTC+2022e}.  However, this rescaling causes SANE simulations to typically have orders of magnitude larger number density than MADs, due to their intrinsically weaker magnetic fields and lower temperatures \citep{EHTC+2021b,EHTC+2022e}.  Consequently, any additional observables sensitive these variables immediately help break degeneracies and distinguish models.

For example, the degeneracies between $n$, $B$, and $\Theta_e$ can be partially resolved with the spectral index, $\alpha \equiv d\log I/d\log \nu$. Spectral index is mainly sensitive to the optical depth $\tau_\nu$ as well as the temperature and magnetic field strength in the combination $B\Theta_e^2$ (to which the critical synchrotron frequency is dependent) \citep[e.g.,][]{Pandya+2016}.  \citet{Ricarte+2022a} show that GRMHD models span a wide range of spectral indices, and that SANE models typically exhibit more negative spectral indices than MADs at a fixed optical depth due to their lower temperatures.  

\begin{figure*}
  \centering
  \includegraphics[width=\textwidth]{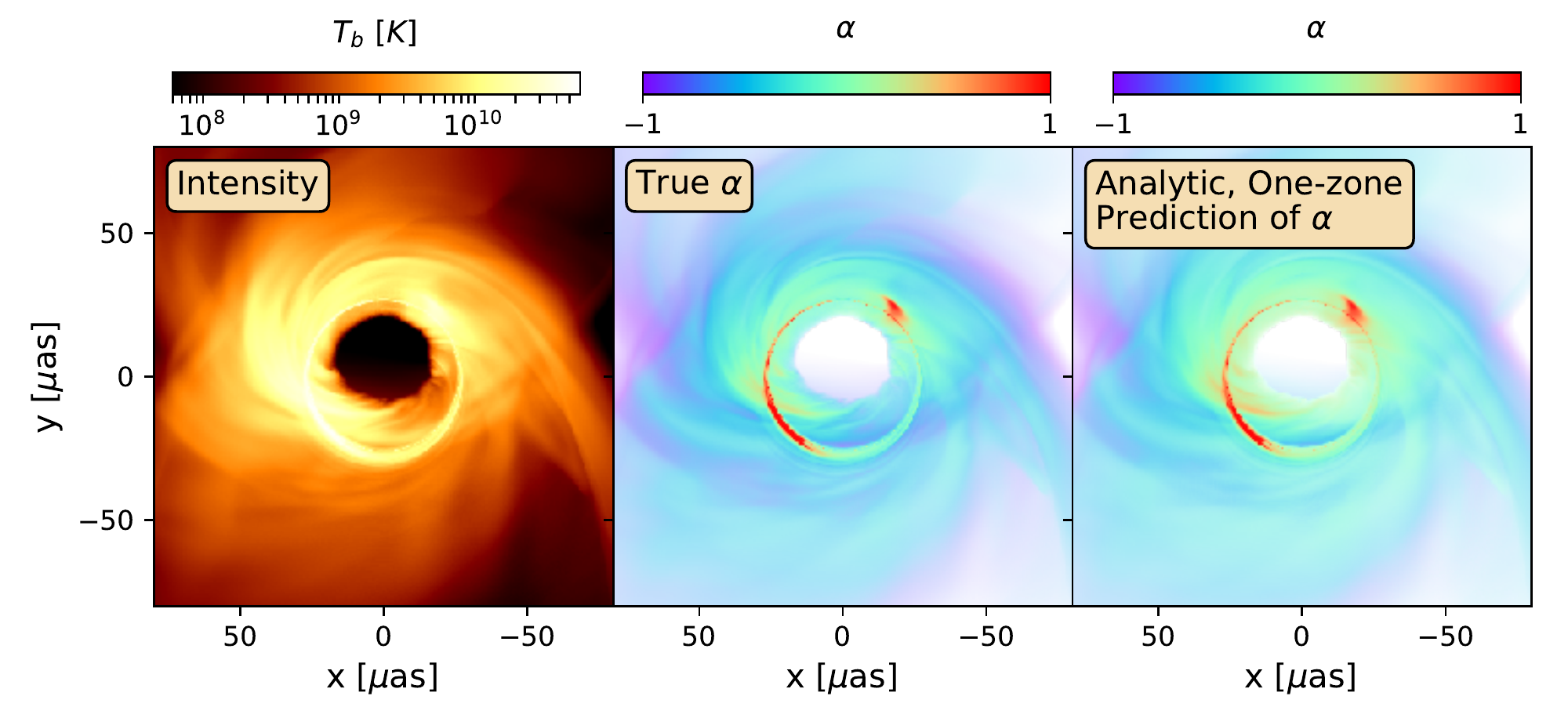}
  \includegraphics[width=\textwidth]{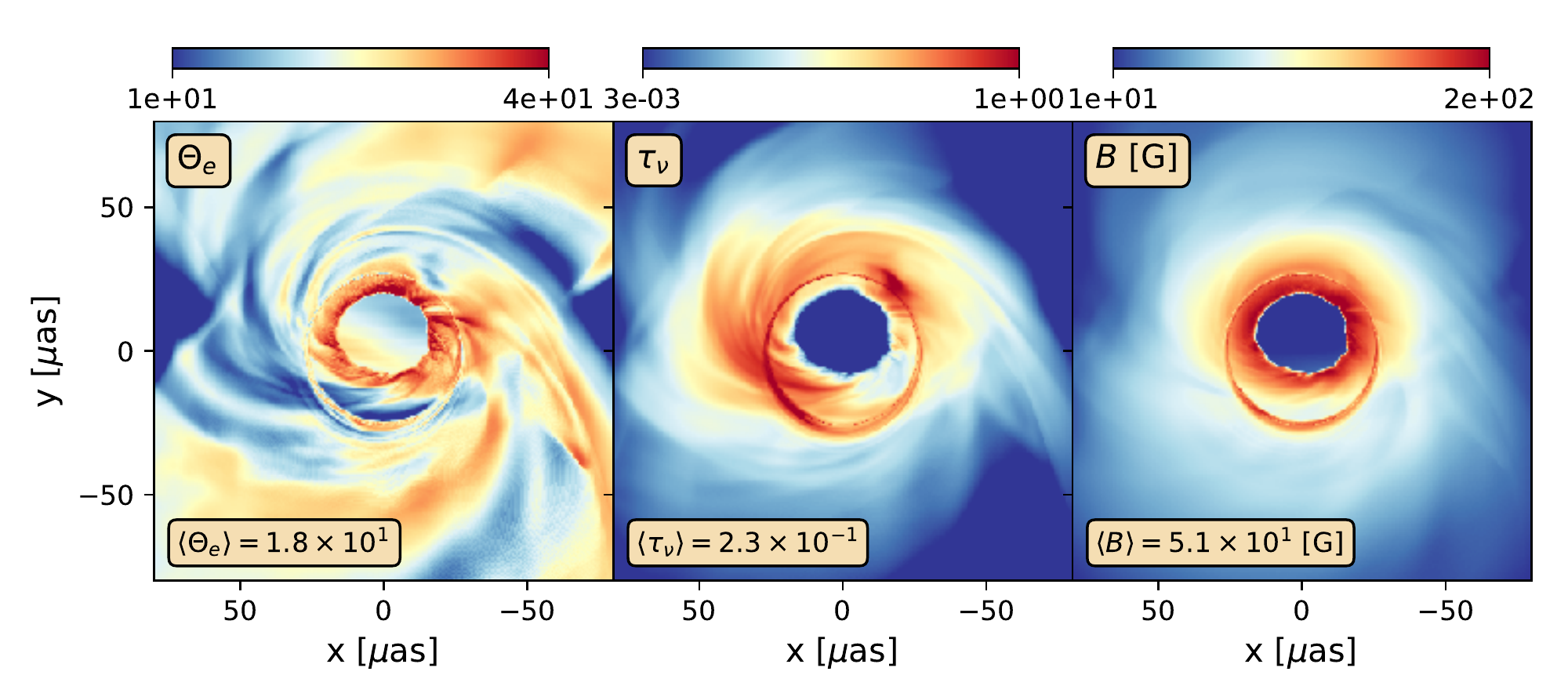}  
  \caption{Intensity and spectral index map of a MAD model of Sgr~A$^*$ \citet[adapted from figure~4 of][]{Ricarte+2022a}.  The top left panel plots total intensity in log scale averaged between 214 and 228 GHz, the top center panel plots the spectral index across this bandwidth calculated by ray tracing the image at two different frequencies, and the top right panel plots an analytic prediction of the spectral index in each pixel obtained by combining the three quantities in the bottom panel: electron temperature, optical depth, and magnetic field strength, each computed by performing an emissivity-weighted average long each geodesic.  The excellent agreement between the true spectral index map and the analytic prediction illustrates the power of spectral index maps to jointly constrain these plasma quantities.}  \label{fig:spectral_index}
\end{figure*}

In the ngEHT era, multi-frequency VLBI will enable not only spatially unresolved spectral index measurements, but also spectral index maps.  Since the most important parameters ($\Theta_e$, $B$, and $\tau_\nu$) all decline with radius, spectral index maps should generically grow more negative as radius increases. Equivalently, the image becomes smaller as the frequency grows larger \citep[e.g.,][]{Blandford&Konigl1979}.  One example from \citet{Ricarte+2022a} is shown in \autoref{fig:spectral_index}, a MAD simulation of Sgr~A$^*$ with $a_\bullet=0$, $R_\mathrm{high}=40$, and a non-thermal ``kappa'' electron distribution function with $\kappa=5$ \citep{Vasyliunas1968,Xiao2006}, inclined at $50^\circ$.  The true spectral index map across 214 to 228 GHz is shown in the top central panel, while a one-zone analytic prediction is shown in the top right, using a $\kappa=5$ eDF combined with the plasma variables computed in the bottom row.  To obtain the analytic prediction, each pixel is treated as a one-zone model using plasma properties computed via an emissivity-weighted average along the geodesic.  In this simulation, both $\tau_\nu$ and $B$ decline with radius, but $\Theta_e$ stays in a relatively narrow range.  Thus, the decline of $\alpha$ with radius can be attributed to a decline in $\tau_\nu$ and $B$.  Models also exhibit a generic spike in $\alpha$ in the photon ring, whose geodesics plunge into regions with strong magnetic fields and acquire a larger optical depth due to their longer path lengths in the emitting region. 

\section{Linear Polarization}
\label{sec:linear}

\begin{figure*}
  \centering
  \includegraphics[width=\textwidth]{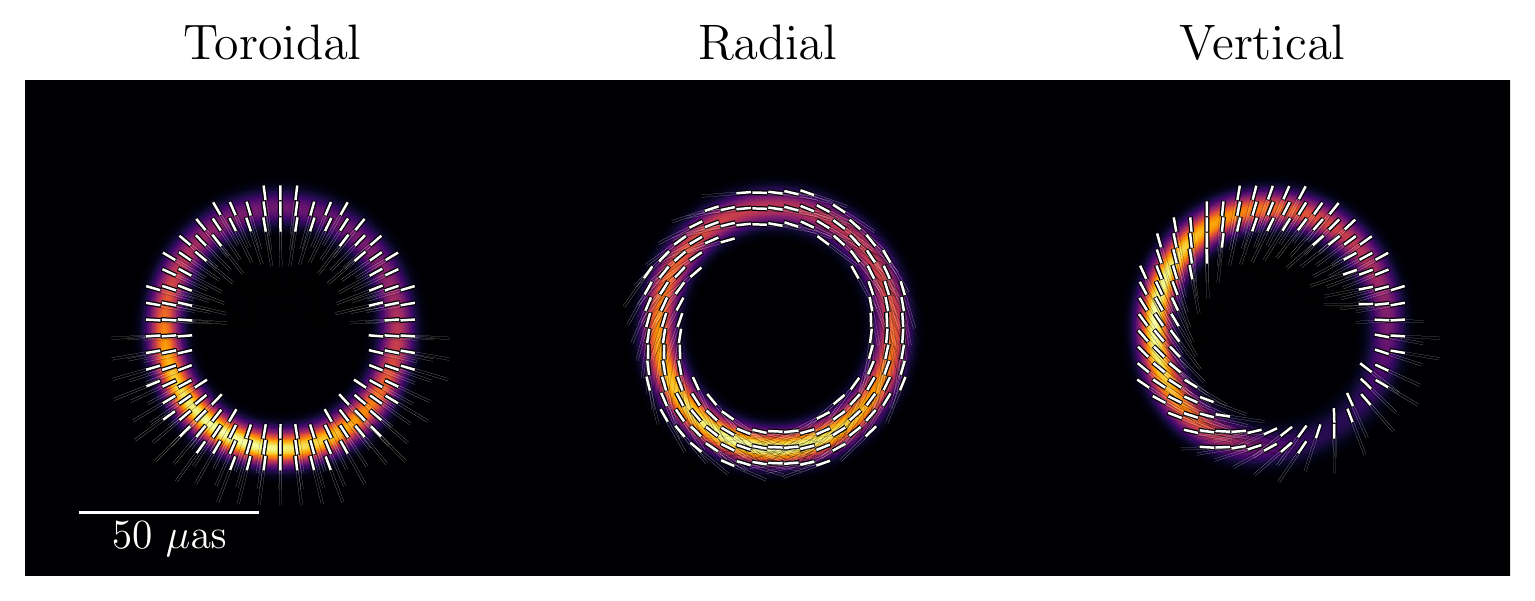}
  \caption{Polarization pattern of a ring of emission around a Schwarzschild black hole threaded with magnetic fields of different geometries: toroidal, radial, and vertical \citep[adapted from Figure~3 of][]{EHTC+2021b}.  The toroidal and radial magnetic field cases clearly illustrate the fact that synchrotron emission is polarized perpendicular to the magnetic field projected onto the sky.  The orientation of the ticks in the vertical field case encodes the direction of the fluid's motion \citep{EHTC+2021b}, chosen here to be clockwise on the sky.  These maps were computed using the analytic ring model of \citet{Narayan+2021}.}  \label{fig:ring_model}
\end{figure*}

A uniform parcel of synchrotron emitting optically thin plasma intrinsically produces a linear polarization fraction of $\approx 70\%$, with an orientation perpendicular to the local magnetic field projected onto the sky \citep[e.g.,][]{Rybicki&Lightman1986}.  The observed orientation of this linear polarization gets modified by two effects: achromatic effects of propagation through the curved space-time, and chromatic ``Faraday'' effects of propagation through a magnetized plasma \citep[recent theoretical investigations in a the vicinity of a black hole include ][]{Himwich+2020,Narayan+2021,Gelles+2021,Palumbo&Wong2022}.  Both rotate and potentially scramble the electric vector position angle (EVPA, or $\chi$).  
Thus, resolved images of linear polarization can inform us about the magnetic field geometry, the magnitude of Faraday effects, and potentially the space-time itself.  In \autoref{fig:ring_model}, we plot the linear polarization pattern of a ring of emission moving clockwise on the sky around a Schwarzschild black hole using the model of \citet{Narayan+2021}.  Here, only the direct ($n=0$) sub-image is included.  The ring is located at a radius of $4.5 \ GM_\bullet/c^2$ and has a Gaussian width of $2 \ GM_\bullet/c^2$.  The angular momentum vector of the ring projected onto the sky has a position angle of $288^\circ$ East of North and is viewed at an inclination of $17^\circ$, consistent with that of M87$^*$ \citep{Walker+2018}.  In the toroidal and radial field cases, the resulting linear polarization pattern is simply perpendicular to the magnetic fields projected onto the line-of-sight.  In the idealized vertical case, the EVPA pattern becomes more sensitive to the direction of the fluid's motion \citep{EHTC+2021b}.

EHT studies have identified the linear polarization fraction (on both resolved and unresolved scales) as well as the morphology of polarization ticks as important observables for theoretical interpretation.  This ``twistiness'' can be quantified by the quantity $\beta_2$, the azimuthally symmetric component of a Fourier decomposition of the polarization pattern \citep{Palumbo+2020}.  M87$^*$ and Sgr~A$^*$ both exhibit much lower linear polarization fractions than the ideal value of 70\% for a uniform parcel of emitting plasma \citep{Goddi+2021}, as do other low luminosity AGN \citep[e.g.,][]{MOJAVE_XV}. \citet{EHTC+2021b} found that such low polarization fractions could be obtained by significant Faraday rotation scrambling.  Combining resolved linear polarization information with an upper limit on circular polarization, \citet{EHTC+2021b} found that MAD models were favored over their SANE counterparts for M87$^*$, which could not be concluded based on total intensity alone.  Fundamentally, this can be attributed to linear polarization's sensitivity to the geometry of the magnetic field, as well as Faraday rotation's sensitivity to cooler electrons that may otherwise be invisible.  MAD models tend to have more ordered fields with stronger poloidal components, which produces twistier polarization patterns.  Meanwhile, SANE models tend to require orders of magnitude larger mass density to compensate for their intrinsically weaker magnetic fields and lower temperatures, resulting in much greater Faraday depths.  In retrograde systems, images can exhibit flips in the sign of $\angle \beta_2$ that correspond to a flip in the angular velocity of inflowing streams due to frame dragging \citep{Ricarte+2022b}.  

Sgr A$^*$ also exhibits interesting time variability in linear polarization, especially during flares, which are accompanied by large polarization fractions, swings in polarization angle, and ``$Q$-$U$ loops'' on the timescale of hours \citep{Marrone+2006,Trippe+2007,Zamaninasab+2011,Gravity+2018,Wielgus+2022b}.  These can be interpreted as the motion of hotspots or other structures as they light up different parts of the magnetic field structure during their orbit \citep{Broderick&Loeb2005,Gravity+2020b,Gelles+2021,Vos+2022}.  The hotspots themselves may originate from ``flux eruption events'' and magnetic reconnection that occur naturally in MAD accretion flows \citep{Ripperda+2020,Dexter+2020,Chatterjee+2022}.  Thus, time variability of linear polarization offers unique insights into the magnetic field structure and direction of orbital motion that could potentially be linked to the inclination and spin of Sgr A$^*$.  The GRAVITY Collaboration has detected centroid motion coincident with a flare \citep{Gravity+2018,Gravity+2020a}.  Spatially resolved movies created by the ngEHT would help test the hotspot interpretation, motivating high-cadence monitoring of this source.

\section{Rotation Measure}
\label{sec:rotation_measure}

In an ionic plasma, circularly polarized waves of opposite handedness propagate at different speeds, resulting in a circular birefringence effect known as Faraday rotation.  The EVPA of propagating emission rotates an amount sensitive to the density, temperature, and line-of-sight magnetic field.  As examined in several studies, internal Faraday rotation is important for depolarizing and scrambling images of GRMHD models of black hole accretion flows \citep{Moscibrodzka+2017,Jimenez-Rosales&Dexter2018,Ricarte+2020,EHTC+2021b}.  The magnitude of Faraday rotation has a wavelength-squared dependence, thus it is useful observationally to define the rotation measure $\mathrm{RM} = d\chi/d\lambda^2$, which offers insights into physical parameters of the Faraday rotating plasma.  For a linearly polarized emitter entirely behind a uniform of Faraday screen, the RM is related to the properties of the screen via

\begin{equation}
    \mathrm{RM} = 8.1\times10^5 \ \mathrm{rad}\; \mathrm{m}^{-2} \int_\mathrm{source}^\mathrm{observer} f_\mathrm{rel}(\Theta_e)\frac{n_e}{1 \ \mathrm{cm}^{-3}} \frac{B_{||}}{G}\frac{ds}{\mathrm{pc}},
    \label{eqn:rm_physics}
\end{equation}

\noindent where $n_e$ is the electron number density, $B_{||}$ is the component of the magnetic field parallel to the photon wave-vector, and $f_\mathrm{rel}$ is a correction term suppressing Faraday rotation at relativistic temperatures \citep{Gardner&Whiteoak1966}.  For relativistic plasmas, $f_\mathrm{rel}(\Theta_e) \approx \log(\Theta_e)/(2\Theta_e^2)$, while $f_\mathrm{rel}$ asymptotes to 1 as $\Theta_e \to 0$.  Here $\Theta_e \equiv k_BT_e/m_ec^2$, $k_B$ is the Boltzmann constant, $T_e$ is the electron temperature, $m_e$ is the electron rest mass, and $c$ is the speed of light \citep{Jones&Odell1977}.  In GRMHD models, the plasma responsible for synchrotron emission is sometimes completely separate from the plasma responsible for Faraday rotation.  For example, some large $R_\mathrm{high}$ SANE models exhibit a cold Faraday rotating midplane sandwiched between emission from their hot jet sheaths \citep{Ricarte+2020}.  Hence, rotation measure and linear polarization can offer a view into electron populations that may otherwise be undetectable from total intensity alone.

\begin{figure*}
  \centering
  \includegraphics[width=\textwidth]{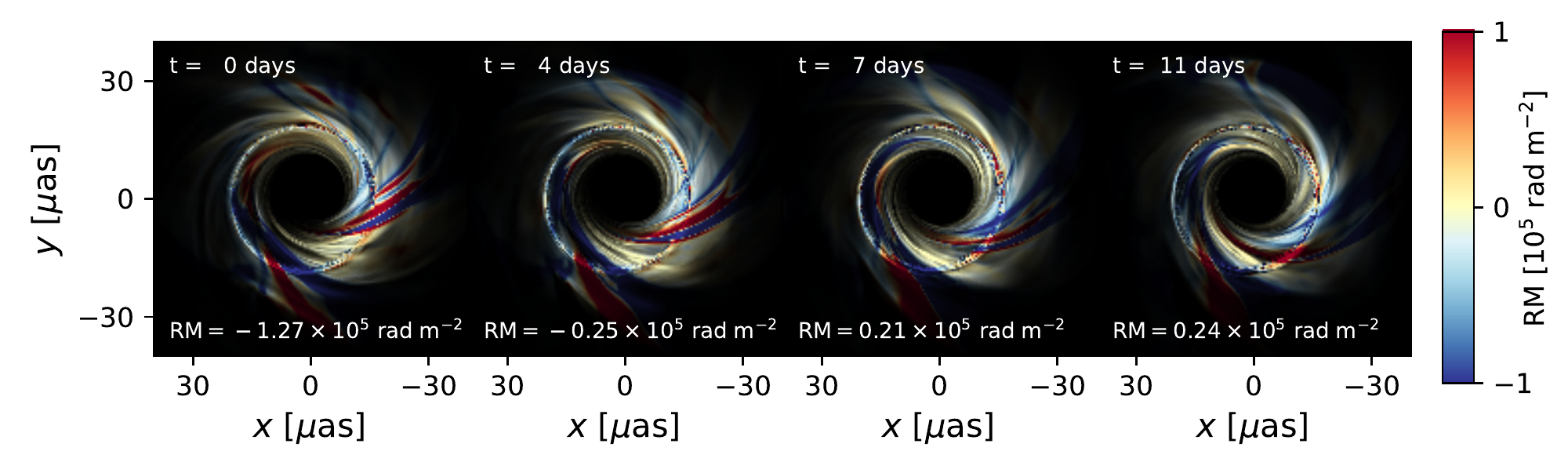}
  \caption{Rotation measure map of a MAD simulation of M87$^*$ \citet[adapted from figure~13 of][]{Ricarte+2020}.  Both positive and negative RM regions are simultaneously present, reflecting flips in the line-of-sight magnetic field direction due to turbulence in the accretion flow.  The motion of these structures produces a time variable spatially unresolved RM, written at the bottom of each panel.}  \label{fig:rotation_measure}
\end{figure*}

At the time of writing, narrow observing bandwidths inhibit our ability to create spatially resolved rotation measure maps, but spatially unresolved RM measurements at millimeter wavelengths exist for Sgr~A*, the core of M87$^*$, and several other LLAGN \citep[e.g.,][]{Marrone_2007,Plambeck_2014,Kuo_2014,Bower_2018,Park_2019,MartiVidal_2019}.  Note also that rotation measure from AGN generally increases with increasing frequency \citep[e.g.,][]{Kravchenko2017} and can reach values of the order of $10^7 \mathrm{rad\,m^{-2}}$ \citep{Ivan2015} due to the opacity effect probing regions close to the central engine at the ngEHT frequencies.  Without spatial resolution, unresolved rotation measure measurements are difficult to interpret because the assumptions underlying \autoref{eqn:rm_physics} are not believed to generally hold.  In GRMHD models, Faraday rotation occurs co-spatially with the plasma, can vary by orders of magnitude in different locations, and can also flip sign across the image due to turbulence \citep{Ricarte+2020}.  As a result, unresolved EVPA measurements may exhibit significant temporal variation and not strictly follow a $\lambda^2$ law.  \autoref{fig:rotation_measure} plots a rotation measure map of a MAD GRMHD model of M87$^*$, with the spatially unresolved RM written at the bottom of each panel.  This turbulence can explain the intra-week time variability of the RM observed for M87$^*$ \citep{Goddi+2021}.  On the other hand, Sgr~A$^*$ has exhibited a constant sign of RM for decades, suggesting the existence of a more stable (but still variable) foreground Faraday screen \citep{Bower+2018}.  Spatially resolved rotation measure maps could help disentangle the Faraday screen and give insights into both the turbulence of the accretion flow and the magnetic field structure of jets. This may be of increased importance, since EHT observations of Sgr~A$^*$ indicate that GRMHD models are too variable \citep{EHTC+2022e}.


\section{Circular Polarization}
\label{sec:circular}

\begin{figure*}
  \centering
  \begin{tabular}{cc}
  \includegraphics[align=c,width=0.33\textwidth]{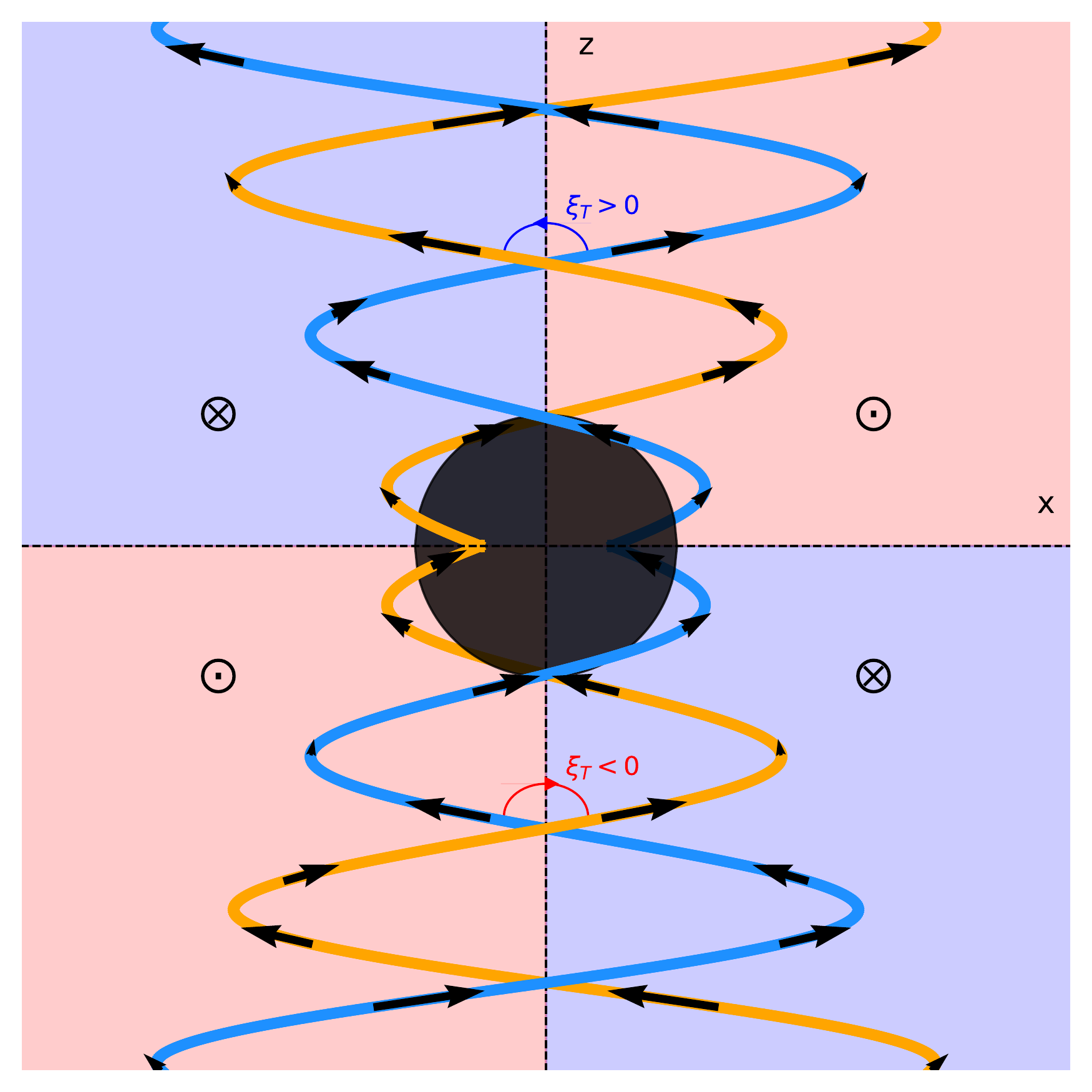} &
  \includegraphics[align=c,width=0.67\textwidth]{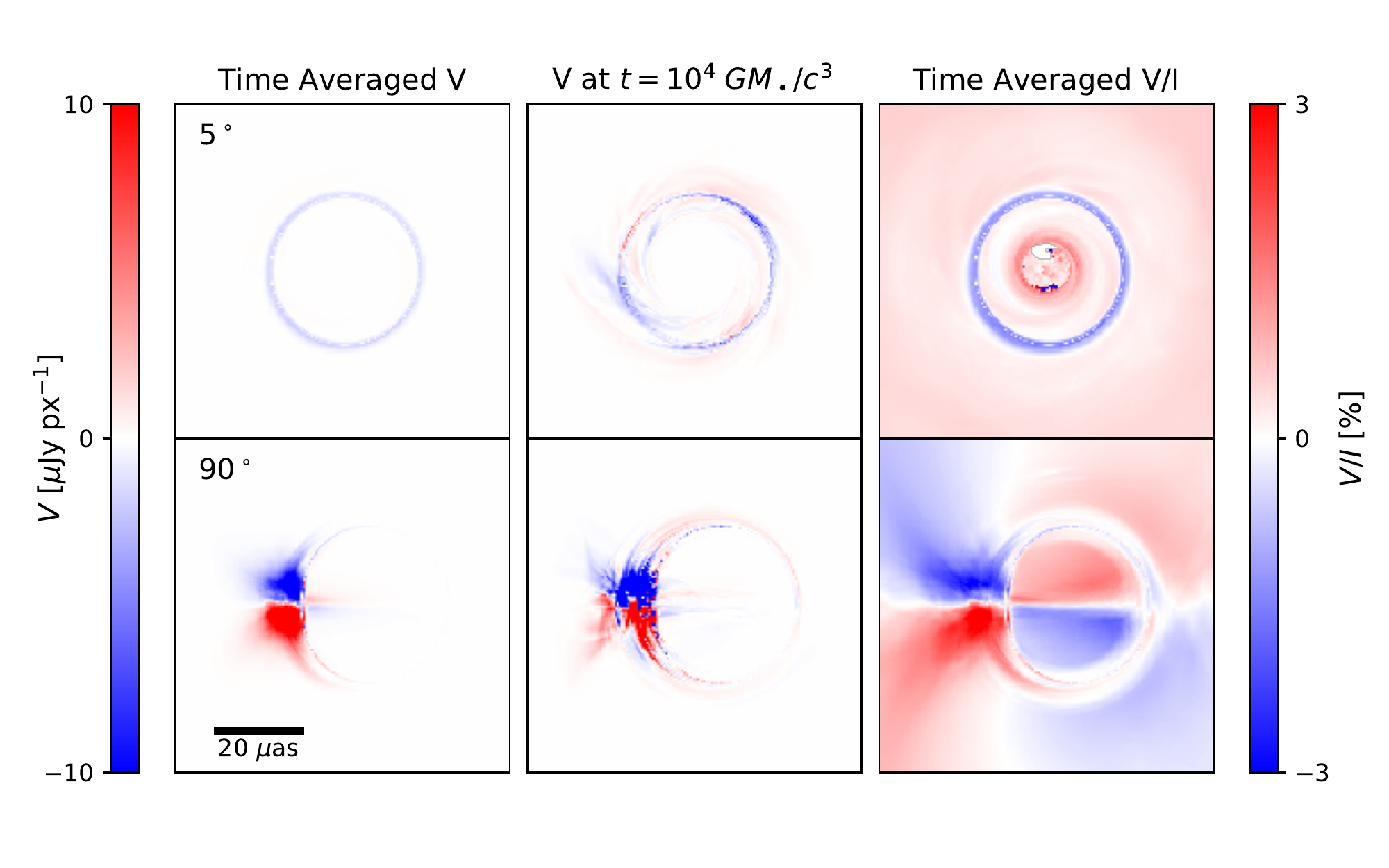}\\
  \end{tabular}
  \caption{Maps of circular polarization encode properties of the geometry of the magnetic field, both its line-of-sight direction and twist.  A cartoon of a generic helical field geometry is depicted on the left.  On the right, we plot the circular polarization of a MAD model of M87$^*$ at two inclinations.  Both are reproduced from \citet{Ricarte+2021}.  The top row depicts a 5$^\circ$ viewing angle, and the bottom row depicts a 90$^\circ$ viewing angle.  The first column shows the time averaged circularly polarized image, the second column shows the same at a single snapshot, and the third column shows fractional circular polarization.  For face-on viewing angles, the photon ring exhibits an interesting sign flip due to Faraday conversion and the sourcing of photons from the opposite side of the disk.  For edge-on viewing angles, circular polarization exhibits a ``four quadrants'' pattern that reflects the line-of-sight magnetic field direction.}  \label{fig:stokes_V}
\end{figure*}

Circular polarization, Stokes V, can be generated both intrinsically through synchrotron emission or through Faraday conversion, which exchanges linear and circular polarization states \citep[e.g.,][]{Wardle&Homan2003}.  Circular polarization fractions are much lower than their linear counterparts for both Sgr~A$^*$ ($V/I \approx -1\%$) and M87$^*$ ($|V/I| \lesssim 0.8\%$), making it more challenging to study than linear polarization.  In addition, the circular feed basis used for EHT sites makes it more challenging to construct circularly polarized images.  However, Stokes~V has the potential to encode not only the magnetic field direction and geometry, but also the plasma composition.

Unlike the near unity linear polarization fractions produced by a uniform parcel of plasma, intrinsically emitted Stokes V is typically produced at the $\sim$1\% level for plasma parameters appropriate for M87$^*$ or Sgr~A$^*$.  The sign of intrinsically emitted circular polarization generated encodes the sign of the magnetic field along the line of sight, following the right hand rule.  The second is Faraday conversion, which exchanges linear and circular polarization states.  The sign of circular polarization generated by conversion depends on the relative orientation of the EVPA with respect to the local magnetic field.  Specifically, Stokes V generated by conversion inherits the sign of Stokes U, defined in the local plasma frame.  Thus, Stokes V from Faraday conversion is sensitive to the line of sight ``twist'' in the magnetic field as well as any Faraday rotation affecting the EVPA of the linear polarization that gets converted.  Interestingly, Faraday conversion has no effect in a unidirectional magnetic field lacking Faraday rotation.

Both intrinsic emission and Faraday conversion are believed to be important for generating circular polarization on event horizon scales \citep{Moscibrodzka+2021,Ricarte+2021,Tsunetoe+2021}.  For Faraday conversion, both the line of sight twist in the magnetic field and Faraday rotation are important for determining the orientation of the linear polarization that is converted into circular.  Small amounts of Faraday rotation can affect the relative alignment between linear polarization and the converting magnetic field.  Large amounts of Faraday rotation can lead to depolarization by randomizing the sign of Stokes U that is converted into Stokes V.

\autoref{fig:stokes_V} depicts a cartoon of a typical helical field geometry as well as the circular polarization produced by a model that exhibits this structure \citep[reproduced from][]{Ricarte+2021}.  In the left panel, the background colors depict the line of sight direction of the magnetic field, viewed edge-on.  In the time averaged image of $V/I$ viewed at 90$^\circ$, This structure produces a ``four quadrants'' pattern at large scales, in the time averaged image of $V/I$ viewed at 90$^\circ$, where on large scales Stokes V originates from intrinsic emission \citep[see also][]{Tsunetoe+2021}.  Another interesting feature that arises due to a generic helical field geometry is the successive sign flipping of sub-images in the photon ring \citep[also discussed in][]{Moscibrodzka+2021}, which can be explained by Faraday conversion and parallel transport in a generic helical field geometry viewed face-on \citep{Ricarte+2021}.  In this particular model, the spatially unresolved Stokes V is surprisingly dominated by this sign-flipped photon ring.

Finally, circular polarization is strongly affected by plasma composition, and can potentially be used to distinguish pair plasmas from ionic plasmas \citep[e.g.,][]{Wardle+1998}.  Most models used to study EHT images have contained only ionic plasma.  Electron-positron pairs can also be naturally produced on event horizon scales, but their abundance is highly theoretically uncertain.  In a pair plasma, intrinsic circularly polarized emission and Faraday rotation both vanish, but Faraday conversion persists.  Intuitively, a pair plasma with equal parts positively and negatively charged particles should not gyrate in a preferred direction.  This can cause dramatic differences in images of circular polarization, and potentially also those of linear polarization \citep{Anantua+2020,Emami+2021}.

\section{Scattering}
\label{sec:scattering}

A major challenge for studies of Sgr~A* with the EHT and ngEHT is interstellar scattering by dilute plasma in the ionized interstellar medium. In particular, the line of sight to Sgr~A* is heavily scattered by plasma in the spiral arms of the Milky Way \citep{Bower_2014}, resulting in angular broadening that is approximately three orders of magnitude larger than median values for lines of sight at higher galactic latitudes. The effects of scattering are two-fold: 1) small-scale modes in the scattering material result in diffractive ``blurring,'' described by a convolution with an anisotropic kernel, and 2) large-scale modes in the scattering material result in refractive ``substructure,'' described by additive image noise with a slowly falling power spectrum. For detailed discussion of the scattering of Sgr~A*, see \citet{Psaltis_2018,Johnson_2018,Johnson_2021}.

While the scattering severely affects images of Sgr~A*, many polarimetric properties of the images are comparatively immune because the scattering is not significantly birefringent. For example, in the case of purely diffractive scattering, the image-integrated fractional polarization is independent of scattering. More generally, the interferometric fractional polarization $\breve{m}(\mathbf{u}) \equiv \tilde{P}(\mathbf{u})/\tilde{I}(\mathbf{u})$ is independent of diffractive scattering because convolution is multiplicative in the visibility domain and is identical for all Stokes parameters, thereby canceling in the quotient. 

\begin{figure*}
  \centering
  \includegraphics[width=0.49\textwidth]{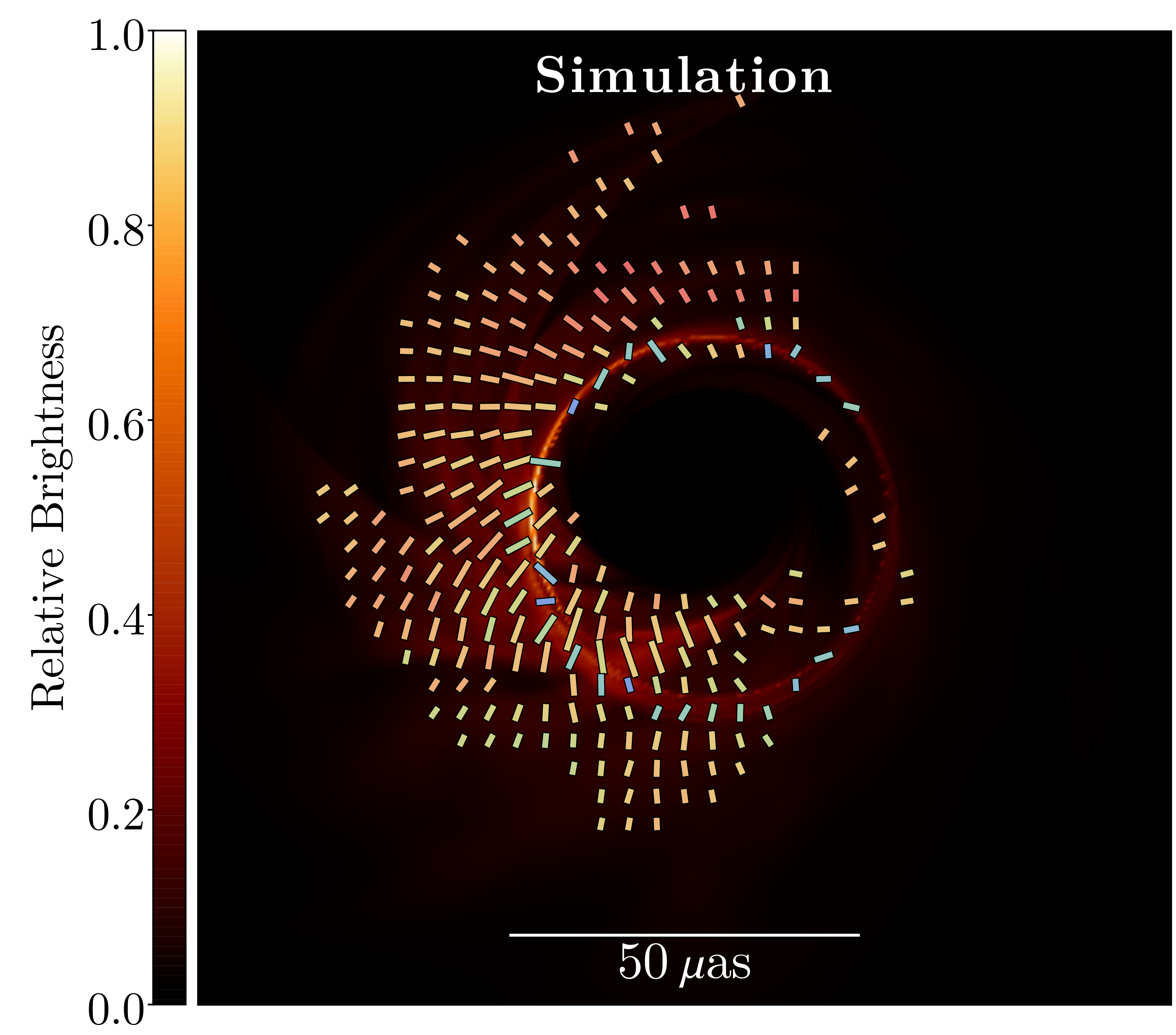}
  \includegraphics[width=0.49\textwidth]{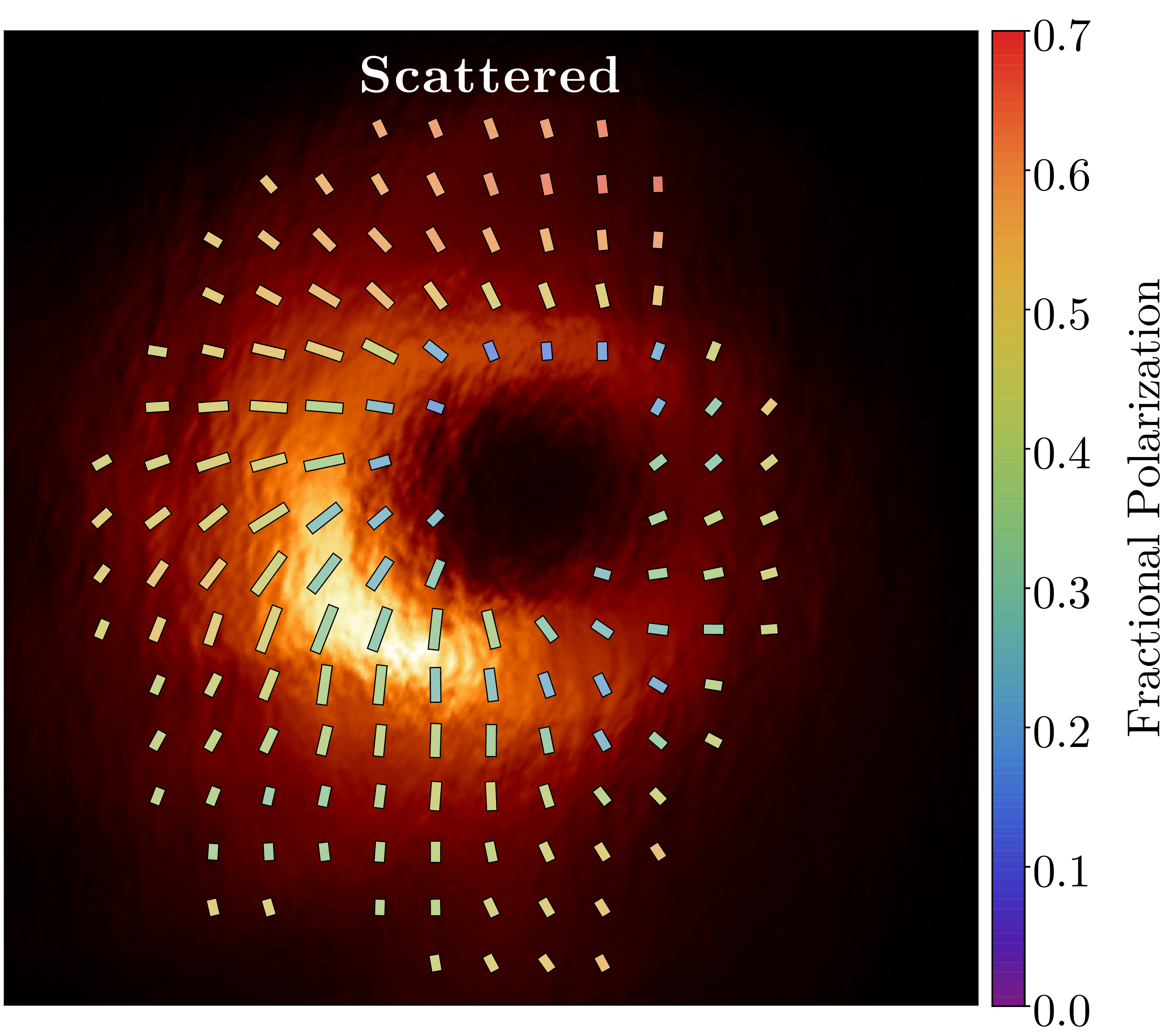}
  \caption{GRMHD model of Sgr~A* at 230\,GHz before (left) and after (right) including the effects of interstellar scattering. This simulation is a MAD with $a_\bullet=0.7$, $R_\mathrm{high}=20$, and $i = 30^\circ$. 
  The background image shows total intensity with respect to the image peak, while the ticks show the polarization magnitude and direction, colored by fractional polarization. While scattering severely affects the image, key polarimetric measures are nearly immune to scattering. For example, the unresolved fractional polarization is 10.5\% before scattering and is 10.6\% after scattering. Likewise, the $\beta_2$ mode in polarization \cite{PWP} has $|\beta_2| = 0.40$ and $\arg(\beta_2) =52.1^\circ$ before scattering, and $|\beta_2| = 0.37$ and $\arg(\beta_2) =51.0^\circ$ after scattering.}  \label{fig:pol_scatt}
\end{figure*}

\begin{figure*}
  \centering
  \includegraphics[width=\textwidth]{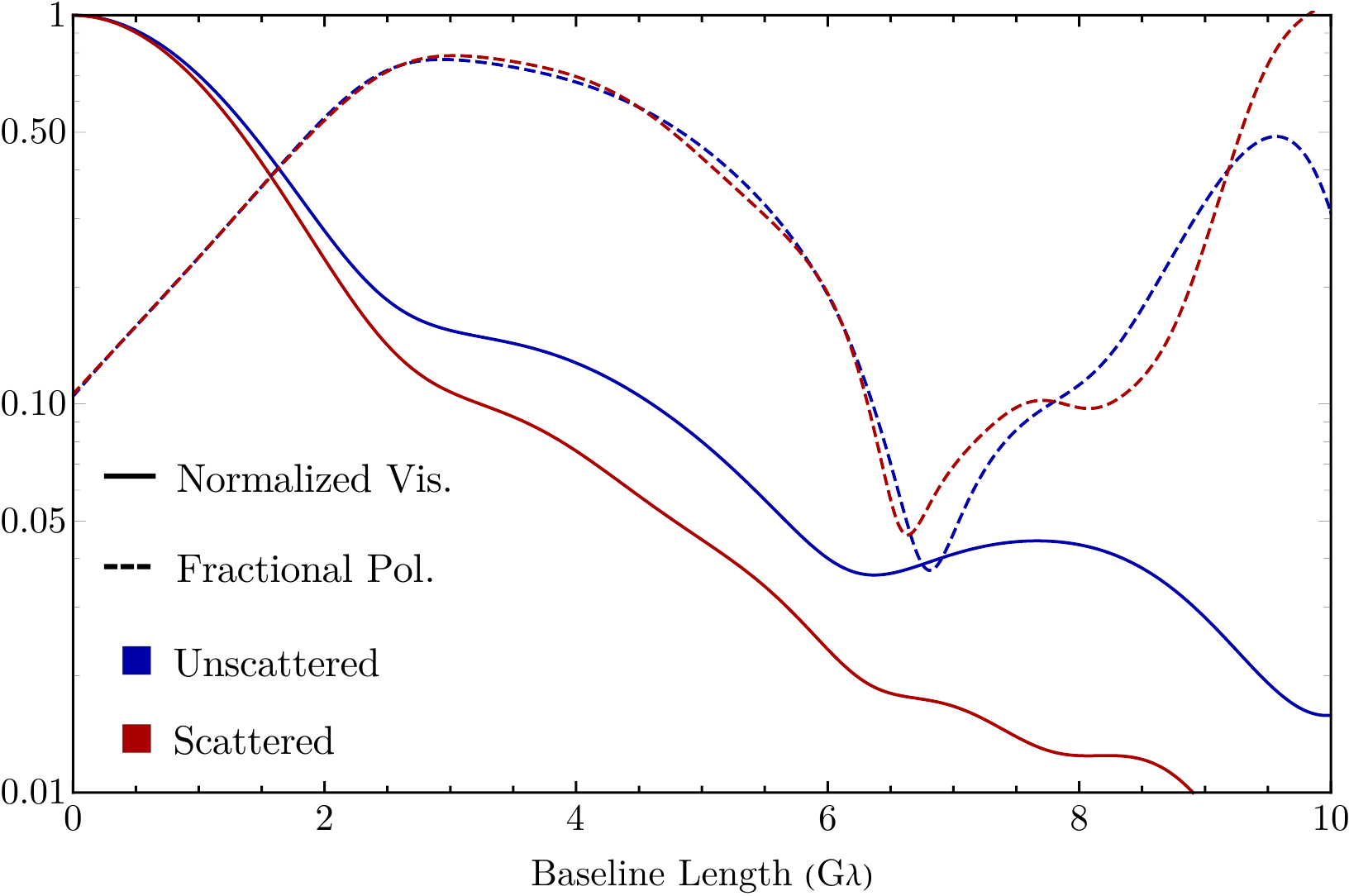}
  \caption{Interferometric properties of the GRMHD model shown in \autoref{fig:pol_scatt}. Solid lines show the normalized intensity $\left| \tilde{I}(\mathbf{u})/\tilde{I}(\mathbf{0}) \right|$ before (blue) and after (red) scattering, and the dashed lines show the interferometric fractional polarization magnitude $\left| \breve{m}(\mathbf{u}) \right|$. For these curves, baselines are oriented along the East-West direction: $\mathbf{u} = (u, 0)$. Over the full range of baseline lengths accessible from the ground, the fractional polarization is largely immune to scattering, while diffractive scattering causes a substantial reduction in the flux on long baselines.}  \label{fig:pol_scatt_vis}
\end{figure*}

\autoref{fig:pol_scatt} shows an example GRMHD snapshot before and after scattering. While the images look substantially different, key polarimetric observables such as the $\beta_2$ mode, which is highly constraining for GRMHD models \cite{EHTC+2021b} and carries information about black hole spin \cite{PWP}, are almost unaffected by scattering. Likewise, certain interferometric observables, such as the interferometric fractional polarization, are only mildly affected by scattering (see \autoref{fig:pol_scatt_vis}). For additional discussion of how the deterministic frequency dependence of scattering can be used for scattering mitigation on images, see \cite{Johnson_2016}; for discussion of how the lack of birefringence can be used to study the relative power spectra in different polarization modes, see Ni et al.~(in prep).

\section{Studying Polarimetry with Interferometry}
\label{sec:interferometry}

Thus far, the discussion has focused on the polarimetric properties of simulated black hole images or images reconstructed from interferometric visibilities. In practice, since the measured visibilities are actually samples of the Fourier transform, image reconstruction can introduce significant systematic uncertainties.  Reconstruction methods must find images consistent with incomplete and noisy information in Fourier space, to which there can be multiple families of solutions.  Images cannot be constructed at all without sufficient $uv$ coverage (or strict image priors).  Thus, it can be useful to study signals in their native visibility space.

\begin{figure*}
  \centering
  \includegraphics[width=\textwidth]{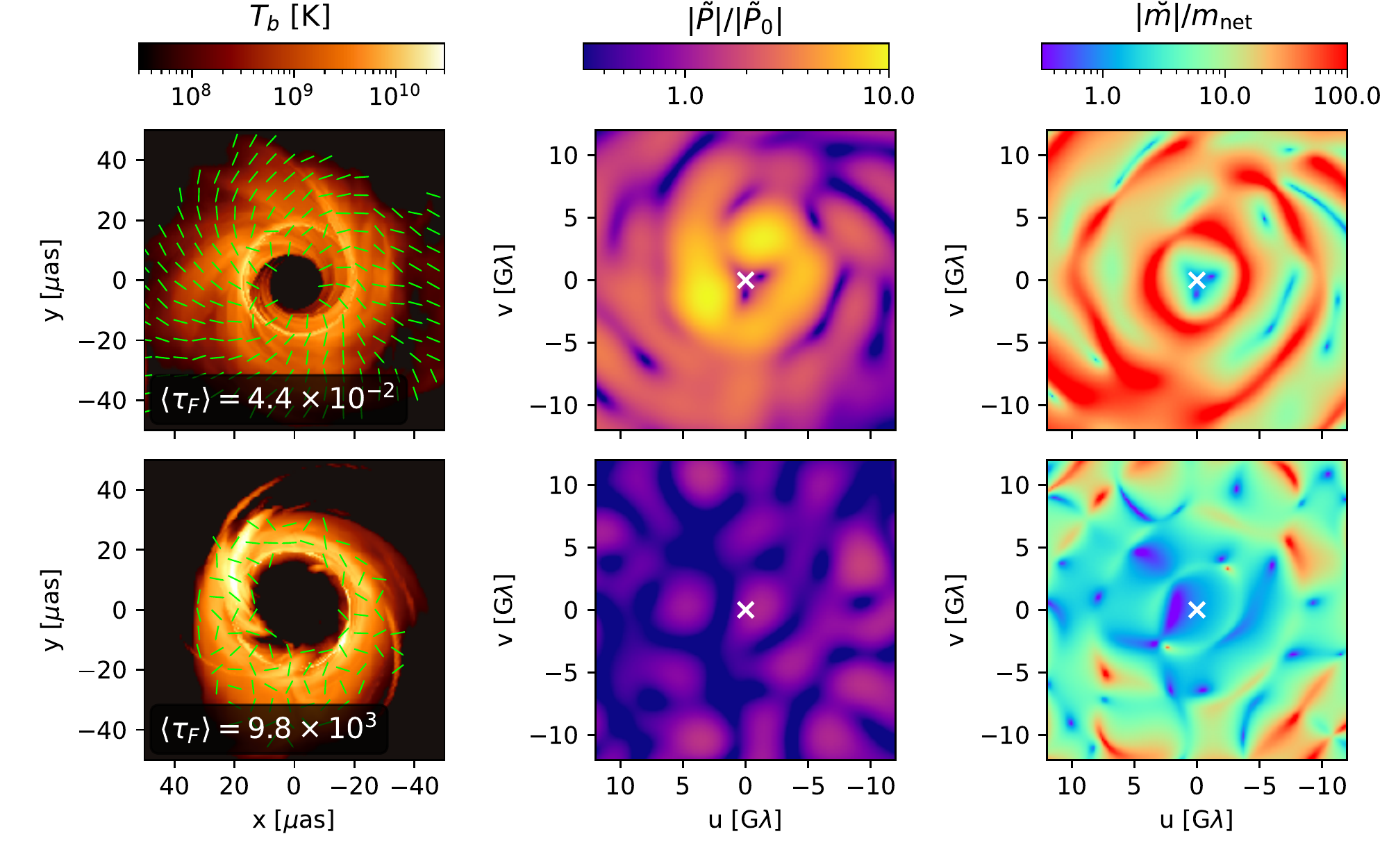}
  \caption{Two GRMHD models imaged at 228 GHz and corresponding maps of linear polarization in visibility space.  The top row corresponds to a MAD model of M87$^*$ with $a_\bullet=0.9$ and $R_\mathrm{high}=1$, while the bottom row corresponds to a SANE model with $a_\bullet=-0.3$ and $R_\mathrm{high}=40$.  Due to a much larger Faraday depth, written at the bottom of the images, the SANE model exhibits a much more disordered linear polarization pattern.  In the ordered model, measures of the linear polarization rise dramatically with radius in the Fourier domain, while the disordered model is characterized by blobs with a coherence length corresponding to the size of the image.}  \label{fig:interferometry}
\end{figure*}

The visibility-domain response of polarimetric observables has some key differences from that of total intensity. For instance, the visibility amplitude for total intensity is guaranteed to be maximal for the zero baseline because the image is positive. However, because the Stokes parameters $Q$, $U$, and $V$ are not constrained to be positive, their visibility functions may not be maximal on the zero baseline. This simple property can be used to make powerful inferences from sparse measurements (e.g., from the EHT or {\it RadioAstron}). For instance, a single measurement of $|\tilde{V}(\mathbf{u})| > |\tilde{V}(\mathbf{0})|$ would demonstrate that the image does not have uniform \emph{sign} of circular polarization. Likewise, if $\left|\tilde{P}(\mathbf{u})\right| \neq \left|\tilde{P}(-\mathbf{u})\right|$, where $\tilde{P}(\mathbf{u}) \equiv \tilde{Q}(\mathbf{u}) + i \tilde{U}(\mathbf{u})$, then the linear polarization field \emph{must} have variations in direction. This test can be performed with as few as two stations, but each must have dual polarization receivers.

A more significant difference between the total intensity and polarization is that the linear polarization can have changes in both amplitude and \emph{direction}, allowing it to have significant image substructure relative to the intensity image. In the visibility domain, this substructure translates to a relative increase in the power on long baselines in polarization versus total intensity. In the limit of a heavily resolved source, the polarimetric signal may exceed that of the total intensity, even for a source with a low fractional polarization!

This can be quantified using the interferometric fractional polarization, $\breve{m}(\mathbf{u}) \equiv \left[ \tilde{Q}(\mathbf{u}) + i \tilde{U}(\mathbf{u}) \right]/\tilde{I}(\mathbf{u})$. This complex quantity corresponds to the unresolved fractional polarization on a zero baseline, $\breve{m}(\mathbf{0}) = (Q_{\rm tot} + i U_{\rm tot})/I_{\rm tot}$, where the ``tot'' subscript denotes an image-integrated quantity. However, unlike the image fractional polarization, $|\breve{m}(\mathbf{u})|$ can exceed unity on long baselines. This was found out from ground-based observations of for Sgr~A* \citep{Johnson2015} as well as within the \textit{RadioAstron} Space VLBI survey of AGN \citep{Kovalev2020}. These observations had one common feature: a very high angular resolution corresponding to several G$\lambda$ spatial scales. In general, for a heavily resolved source with polarized substructure, we expect $|m(\mathbf{u})|$ to generically grow with increasing baseline length. 

From a calibration perspective, the interferometric fractional polarization has the benefit of properties analogous to VLBI closure quantities, since the source of rapidly varying gains at VLBI sites (e.g., changing atmospheric delay and reference frequency errors) are equivalent for both polarization feeds. In addition, the interferometric fractional polarization is resilient to the effects of interstellar scattering, which is likewise not significantly birefringent (see \autoref{fig:pol_scatt_vis}). Finally, $|m(\mathbf{u})|$ is a useful observable to measure the relative coherence of the polarization field. For a perfectly uniform polarization field, $|m(\mathbf{u})|$ will be independent of baseline length. However, for a disordered polarization field, $|m(\mathbf{u})|$ will grow roughly as $1/|\tilde{I}(\mathbf{u})|$, as the observations resolve the structure in total intensity without resolving the structure in polarization. 

In \autoref{fig:interferometry}, we plot total intensity and linear polarization maps of two models of M87$^*$ with very different polarization characteristics.  The top model is a MAD simulation with $a_\bullet=0.9$ and $R_\mathrm{high}=1$, which exhibits an ordered polarization pattern due to ordered magnetic fields and little Faraday rotation.  Meanwhile, the bottom model is a SANE simulation with $a_\bullet=-0.3$ and $R_\mathrm{high}=40$.  As mentioned in \autoref{sec:rotation_measure}, SANE simulations tend to have much larger Faraday depths than MADs, causing this model's linear polarization to be much more disordered.  The pixel-to-pixel intensity weighted Faraday depth $\langle \tau_F \rangle$ is written at the bottom of each panel.

These characteristics are reflected in their Fourier space maps of $|\tilde{P}|$ and $|\breve{m}|$, shown in the second and third columns, which can be directly sampled using an interferometer.  Here, $\tilde{P} \equiv \tilde{Q} + i\tilde{U}$ and $\breve{m} \equiv \tilde{P}/\tilde{I}$, where $\sim$ denotes a Fourier transform.  For the MAD model, $\tilde{P}$ and $\breve{m}$ both rise dramatically with radius in Fourier space.  $\tilde{P}$ rises because the linear polarization is higher on resolved scales than a spatially unresolved measurement would suggest.  The rotational symmetry of the polarization pattern causes substantial cancellation of polarization without spatial resolution.  $\breve{m}$ also rises for this reason, and also because $\tilde{I}$ exhibits nulls in Fourier space that do not necessarily coincide with the nulls in $\tilde{P}$.  Meanwhile, the disordered SANE simulation exhibits a mottled pattern in $\tilde{P}$ with a characteristic length scale corresponding to the size of the image.  These phenomena should not change much qualitatively as a function of wavelength in the sub-millimeter.

The previous discussion has focused on the relationship between expected image features and their appearance in the (Fourier-conjugate) interferometric visibility domain. However, a crucial consideration is how to study frequency-dependent effects, such as spectral index and rotation measure, using interferometry. Because the fringe spacing $u \propto \nu$, interferometric measurements across multiple frequencies necessarily mix the effects of a changing dimensionless baseline with those of a changing image. 

Specifically, the interferometric response $V$ on a physical vector baseline $\mathbf{b}$ at an observing frequency $\nu$ is
\begin{align}
    V(\mathbf{b},\nu) &= \int d^2 \boldsymbol{\theta}\,I(\boldsymbol{\theta}, \nu) e^{-2\pi i \frac{\nu}{c} \boldsymbol{\theta} \cdot \mathbf{b}}\\
    \Rightarrow \nu \partial_\nu V(\mathbf{b},\nu) &= 
        \int d^2 \boldsymbol{\theta}\, \left[ 
        \nu \partial_\nu I(\boldsymbol{\theta}, \nu) - 2\pi i \boldsymbol{\theta} \cdot \mathbf{u}  I(\boldsymbol{\theta}, \nu) 
        \right] e^{-2\pi i \boldsymbol{\theta} \cdot \mathbf{u}}.
\end{align}
The first term in the square brackets accounts for the frequency dependence of the image, while the second accounts for the changing dimensionless baseline with frequency. For the first term, an image with spectral index $\alpha$ has $|\nu \partial_\nu I| \sim \alpha I$, while the effects of RM give $|\nu \partial_\nu I| \sim 4\times {\rm RM} \times \lambda^2 I$. Roughly speaking, we expect that the relative dominance or subdominance of spectral index versus RM are independent of baseline length, so the relative effects on long baselines are likely similar to those for unresolved measurements of a source. For instance, the effects of rotation measure for observations of Sgr~A$^*$ at millimeter wavelengths are likely to vastly dominate the effects of spectral index. Sgr~A$^*$ has $\alpha = 0.0 \pm 0.1$ \citep{Wielgus+2022a} but $4\times {\rm RM} \times \lambda^2 \approx -2.7$ \citep{Goddi+2021}.  The second term gives a relative contribution that increases as the image is increasingly resolved. It becomes significant when the spanned frequencies change the baseline length by the inverse field-of-view, $F$. 

\section{Discussion and Conclusion}
\label{sec:conclusion}

The EHT and upcoming ngEHT enable us to probe accreting supermassive black holes via a variety of multi-frequency polarimetric observables.  In this contribution, we have discussed the many ways in which the physical properties of underlying accretion flow are mapped onto these observables.  Total intensity and spectral index encode the density, temperature, and magnetic field strength of emitting plasma in different regions.  Linear polarization encodes the geometry of the magnetic field, and its depolarization via Faraday rotation offers an observational probe into otherwise invisible cool electrons.  Rotation measure maps probe this cooler Faraday rotating electron population directly, and can probe the magnetic field direction, which can reveal turbulent structures.  Finally, circular polarization encodes both overall geometry and direction of the magnetic field via emission, Faraday rotation, and Faraday conversion.  We have discussed that even if imaging proves prohibitively challenging for some datasets, constraining information exists already in visibility space.  For some models, low polarization fractions in spatially unresolved measurements hide large polarization fractions in spatially resolved measurements.  In the ngEHT era, we will have access to not only single snapshots, but also movies, with a much larger dynamic range in intensity than is presently possible with the EHT.  Multiple snapshots will also enable cleaner theoretical connections via time averaging \citep[e.g.,][]{Gold+2017,Pushkarev+2022}.  This will enable an unprecedented deluge of data about the nearest SMBHs that will help us understand their accretion and feedback processes.

\section{Acknowledgments}

We thank the National Science Foundation (AST-1716536, AST-1935980, AST-2034306, AST-1816420, and OISE-1743747) for financial support of this work. 
YYK was supported by the Russian Science Foundation grant 21-12-00241.
This work was supported in part by the Black Hole Initiative, which is funded by grants from the John Templeton Foundation and the Gordon and Betty Moore Foundation to Harvard University.
The opinions expressed in this publication are those of the author(s) and do not necessarily reflect the views of the Moore or Templeton Foundations.

\bibliography{ms.bib}

\end{document}